\documentclass{article}

\usepackage{arxiv}

\usepackage[utf8]{inputenc} 
\usepackage[T1]{fontenc}    
\usepackage{hyperref}       
\usepackage{url}            
\usepackage{booktabs}       
\usepackage{amsfonts}       
\usepackage{nicefrac}       
\usepackage{microtype}      

\usepackage{graphicx}

\usepackage{colortbl}		
\usepackage{multicol}        

\usepackage{amsbsy}
\usepackage{amsfonts}
\usepackage{amsmath}

\providecommand{\bo}{\mathbf}
\providecommand{\bm}{\mathbf}
\providecommand{\bs}{\boldsymbol}
\providecommand{\cov}{\mathrm{\bo COV}}
\providecommand{\E}{\mathrm{\bo E}}

\providecommand{\diag}{\mathrm{diag}}

\newcommand{\indep}{\bot\!\!\,\!\!\bot}

\newtheorem{assumption}{Assumption}
\newtheorem{result}{Result}
\newtheorem{definition}{Definition}

\title{Sliced Inverse Regression for Spatial Data}

\date{} 					

\author{Christoph~Muehlmann \\
	Institute of Statistics \& Mathematical Methods in Economics \\
	Vienna University of Technology \\
	\texttt{christoph.muehlmann@tuwien.ac.at} \\
	\And
	Hannu~Oja \\
	Department of Mathematics and Statistics \\ 
	University of Turku \\
	\texttt{hannu.oja@utu.fi} \\
	\And
	Klaus~Nordhausen \\
	Institute of Statistics \& Mathematical Methods in Economics \\
	Vienna University of Technology \\
	\texttt{klaus.nordhausen@tuwien.ac.at} \\
}

\renewcommand{\headeright}{}
\renewcommand{\undertitle}{}

\begin{document}
\maketitle

\begin{abstract}
Sliced inverse regression is one of the most popular sufficient dimension reduction methods. Originally, it was designed for independent and identically distributed data and recently extend to the case of serially and spatially dependent data. In this work we extend it to the case of spatially dependent data where the response might depend also on neighbouring covariates when the observations are taken on a grid-like structure as it is often the case in econometric spatial regression applications. We suggest guidelines on how to decide upon the dimension of the subspace of interest and also which spatial lag might be of interest when modeling the response. These guidelines are supported by a conducted simulation study. 
\end{abstract}

\section{Introduction}
\label{Muehlmann:sec:1}

Data that is recorded at spatial locations is increasingly common. For such spatial data it is natural to assume that measurements which are close to each other are more similar than measurements taken far apart. In a regression context is also natural to consider that the response is not only a function of the explaining variables measured at the same location but might also depend on explaining variables in the vicinity. For example, if the response is some measurement of pollution at a given location it might depend also on environmental factors in neighbouring areas which could be carried over by the wind. Or, if one is interested in the value of houses in one district the value of the houses in the neighbourhood as well as their crime statistics might be of relevance.

There are many spatial regression models taking into consideration spatial proximity, see for example \cite{IntroductionToSpatialEconometrics,SpatialEconometrics} and references therein. The amount of possible explaining variables measured is however increasing tremendously, hence, it would be beneficial to identify a sufficient lower dimensional subspace of the data prior building the actual regression model. Reducing the dimension of the explaining variables without loosing information on the response is known as sufficient dimension reduction (SDR), it is well established for iid data (see \cite{MaZhu2013} for a recent review). The most popular SDR methods are sliced inverse regression (SIR) \cite{Li1991} and 
sliced average variance estimation (SAVE) \cite{CookWeisberg1991,Cook2000}. SIR and SAVE have been recently extend to the time series case \cite{MatilainenCrouxNordhausenOja2017,MatilainenCrouxNordhausenOja2019}. SDR methods for spatial data are not yet much investigated. \cite{GuanWang2010} suggested several SDR methods for spatial point processes, which means that the locations by itself are stochastic and of interest and thus considered as a response to be modelled using explaining covariates. Another type of spatial data is often referred to as geostatistical data where at fixed locations random phenomena are observed and these should be modelled. In general for geostatistical data locations are usually irregularly selected from the domain of interested. A special case is when the locations are lying on a regular grid.
For such grid data \cite{LoubesYao2009,2019arXiv190909996A} considered kernel SIR and SAVE methods under the assumption that the response at a location is a function of the covariates at this same location.
In this paper we extend SIR to grid data where we assume that the response at a given location might also depend on covariantes measured at different locations. Our approach follows ideas from blind source separation and is an extension of the time series SIR method suggested in \cite{MatilainenCrouxNordhausenOja2017}.


The structure of the paper is as follows, in Chapter~\ref{sec::sir_iid} we recall SIR in a blind source separation framework for iid data. Chapter~\ref{sec::sir_time} is devoted to SIR for time series data as our extension is build on these ideas. Then, in Chapter~\ref{sec::sir_spatial} we suggest our extension of these methods for spatial data. Lastly, we present several simulation studies of our spatial SIR method in Chapter~\ref{sec::sim}.

\section{SIR for iid data}\label{sec::sir_iid}
For the purpose of this paper we follow the notions as in \cite{MatilainenCrouxNordhausenOja2017, MatilainenCrouxNordhausenOja2019} and introduce SIR in a blind source separation model context in which it is assumed that the response $y$ is univariate and the $p$-variate vector $\bo x$ of explaining variables has the representation
\[
\bo x = \bs \Omega \bo z + \bs \mu = \bs \Omega 
 \begin{pmatrix}
    \bo z^{(1)} \\
    \bo z^{(2)}
  \end{pmatrix}
  + \bs \mu,
\] 
where $\bs \mu$ is a $p$-variate location vector and the $p \times p$ matrix $\bs \Omega$ is called the mixing matrix  having the only restriction to be full rank. Regarding the latent unobservable $p$-variate random vector $\bo z$ the following assumptions are made:

\begin{assumption}
The random vector $\bo z$ can be partitioned into the $d$-variate subvector $\bo z^{(1)}$
and the $p-d$-variate subvector $\bo z^{(2)}$ and together they satisfy
\begin{enumerate}
	\item[(A1)] $ \E(\bo z) = \bo 0$ and $ \cov( \bo z) = \bo I_p$, and
	\item[(A2)] $ \left(y, \bo z^{(1)}{}^\top \right)^\top \indep \ \bo z^{(2)}$.
\end{enumerate}
The dimension $d$ and the partitioning  are minimal in the sense that, 
for projection matrices satisfying  $(y,\bo P \bo z)\indep (\bo I- \bo P)\bo z$,
the rank of $\bo P$ is larger than or equal to $d$ and 
$\bo P\left({\bo z^{(1)}}^\top,\bo 0^\top\right)^\top=\left({\bo z^{(1)}}^\top,\bo 0^\top\right)^\top$.
\end{assumption}

Note however that these assumptions do not specify $\bo z$ completely, both subvectors are defined only upto rotation by an orthogonal matrix of corresponding dimension.

Assumption (A2) is slightly different than the one usually stated in the SIR literature where it is required that
\begin{enumerate}
	\item[(A2')] $ \bo z^{(2)} \indep y | \bo z^{(1)}$ and $E(\bo z^{(2)} | \bo z^{(1)}) = \bo 0$ (a.s.).
\end{enumerate}

Thus in (A2') some weak dependence between $\bo z^{(2)}$ and $\bo z^{(1)}$ is possible. To extend the theory to stochastic processes like time series or random fields we make the stronger assumption (A2) which makes it easier to apply tools from blind source separation methods where independence assumptions are frequent. This stronger assumption was also used in the iid case in \cite{Nordhausen:2016} to construct asymptotic and bootstrap tests for $d$, something we also plan to extend in future work to the depended settings described in the subsequent sections.\\

Naturally (A2) implies (A2') and both also have as consequence that
\[
\cov(\E(\bo z| y)) = 
 \begin{pmatrix}
 \cov(\E(\bo z^{(1)}| y)) & \bo 0\\
 \bo 0 & \bo 0
  \end{pmatrix}.
\]

This matrix is the key to SIR which is defined as:
\begin{definition} \label{def:iid:sir}
The sliced inverse regression functional $\bs \Gamma(\bo x, y)$ at the joint distribution of $\bo x$ and $y$ is obtained as follows
\begin{enumerate}
 \item Whiten the explaining vector: $\bo x^{st} = \cov(\bo x)^{-1/2}(\bo x - \E(\bo x))$.
 \item Find the $d \times p$ matrix $\bo U$ with orthonormal rows $\bo u_1, \ldots, \bo u_d$ which maximizes
 \[
 || \diag\left(\bo U \cov(\E(\bo x^{st}|y)) \bo U^\top \right) ||^2 = \sum_{c=1}^d \left(\bo u_c^\top \cov(\E(\bo x^{st}|y)) \bo u_c \right)^2.
 \]
 \item $\bs \Gamma(\bo x, y) = \bo U \cov(\bo x)^{-1/2}$.
\end{enumerate}
\end{definition}

The question is then naturally how to estimate $\cov(\E(\bo x^{st}|y))$ if $y$ is not discrete; that is where the term sliced comes into play. In that case $y$ is discretized and ``sliced'' into $H$ disjoint intervals yielding $y^{sl}$ and then in Definition~\ref{def:iid:sir} the sliced $y^{sl}$ is used rather than $y$ itself. It was shown that SIR is quite robust with  respect to the slicing and usually $H=10$ slices 
are used.

The optimization problem from Definition~\ref{def:iid:sir} can be solved by performing an eigenvector decomposition of $\cov(\E(\bo x^{st}|y))$ or $\cov(\E(\bo x^{st}|y^{sl}))$ respectively. The vectors $\bo u_c$ consist of the eigenvectors of this decomposition which have non-zero eigenvalues. Note that, given the slicing does not cause loss of information, there should be exactly $d$ non-zero eigenvalues and the magnitude of these eigenvalues reflects the relevance of the corresponding direction for the response. 

Hence, when plugging in data from a sample to obtain an estimate of the SIR functional the eigenvalues can give an idea about the usually unknown value of $d$. Inferential theory for the order determination based on these eigenvalues is for example mentioned in \cite{BuraCook2001, Nordhausen:2016,LuoLi2016} and references therein.

 \section{SIR for time series data}\label{sec::sir_time}

As the next step we introduce SIR in the time series context following \cite{MatilainenCrouxNordhausenOja2017} as we will extend this approach to spatially dependent data. Thus, we consider an univariate response time series $y=y[t]$, $t=0, \pm 1, \pm 2, \ldots$ and a $p$-variate times series $\bo x=\bo x[t]$, $t=0, \pm 1, \pm 2, \ldots$, that is used to explain $y$. Similarly as in the spatial data case it is natural to assume that the dependence of $y$ and $\bo x$ might also lag in time and therefore the time structure is taken into consideration when performing the dimension reduction. \cite{BeckerFried2003} suggested to simply add the lag-shifted times series as new variables to the process $\bo x[t]$ yielding $\bo x^*[t] = (\bm x[t]^\top, \bm x[t-1]^\top, \dots, \bm x[t-K]^\top)^\top$ and apply the iid SIR to the pair $y[t]$ and $\bo x^*[t]$. The disadvantage of this approach is that if $K$ lags are of interest then the dimension of $\bo x^*[t]$ is $(K+1)p$ while at the same time the sample size is reduced by $K$.

Another approach for a time series version of SIR was suggested recently by \cite{MatilainenCrouxNordhausenOja2017}, the main idea is to incorporate serial information in   $\cov(\E(\bo x|y))$ by defining 
\[
\Sigma_\tau(\bo x) = \cov(\E(\bo x[t]|y[t + \tau])).
\]

For their method to work \cite{MatilainenCrouxNordhausenOja2017} formulate the time series SIR blind source separation model 
\[
\bo x[t] = \bs \Omega \bo z[t] + \bs \mu = \bs \Omega 
 \begin{pmatrix}
    \bo z^{(1)}[t] \\
    \bo z^{(2)}[t]
  \end{pmatrix}
  + \bs \mu,
\] 
where $\bs \Omega$ and $\bs \mu$ are the full-rank $p \times p$ mixing matrix and the $p$-variate location vector respectively. For the latent unobservable $p$-variate random process $\bo z =\bo z[t] = (\bo z^{(1)}[t], \bo z^{(2)}[t] )^\top$ the following assumptions are made:

\begin{assumption}
The stationary random process $\bo z = \bo z[t]$ can be partitioned into the $d$-variate subprocess $\bo z^{(1)} = \bo z^{(1)}[t]$
and the $p-d$-variate subprocess $\bo z^{(2)} =\bo z^{(2)}[t]$ and together they satisfy
\begin{enumerate}
	\item[(A3)] $ \E(\bo z) = \bo 0$ and $ \cov( \bo z) = \bo I_p$, and
	\item[(A4)] $ \left(y, \bo z^{(1)}{}^\top \right)^\top \indep \ \bo z^{(2)}$,
\end{enumerate}
where $d$ is minimal in the sense as specified in Assumption 1.
\end{assumption}

Thus, all necessary information required to model $y[t]$ goes through $\bo z^{(1)}[t]$ as Assumption (A4) implies also that 
\[ 
\left(y[t_1], \bo z^{(1)}[t_1]^\top \right)^\top \indep \ \bo z^{(2)}[t_2] \quad \mbox{or} \quad \left(y[t_1 + \tau], \bo z^{(1)}[t_1]^\top \right)^\top \indep \ \bo z^{(2)}[t_2]
\]
for all $t_1, t_2, \tau \in \mathbb{Z}$.\\

It holds then that
\[
\Sigma_\tau(\bo z)
	 = \left(
		 \begin{array}{cc}
		 \cov \left( \E( \bm z^{(1)}[t]|y[t+\tau]) \right) & 0 \\
		0 & 0 \\
		 \end{array}
	 \right)
\ \ \mbox{for all $\tau \in \mathbb{Z}.$}
\]

The idea of \cite{MatilainenCrouxNordhausenOja2017} is to jointly diagonalize several matrices $\Sigma_\tau(\bo z)$ using a set of different lags $\mathcal{T}=\{\tau_1, \ldots, \tau_K\}$ rather than diagonalizing only one matrix $\Sigma_\tau(\bo z)$. This leads to the following definition of the time series (TSIR) functional.

\begin{definition}
The TSIR functional ${\bs \Gamma}( \bo x; y)$ for a stationary time series $(y, \bo x^\top)^\top$ is obtained as follows.
\begin{enumerate}
\item Standardize $ \bo x$ and write ${\bo x}^{st} : = {\cov( \bo x)}^{-1/2}({\bo x} - \E( \bo x))$.
\item Find the $d \times p$ matrix $ \bo U = ( \bo u_1, \ldots, \bo u_d)^\top$ with orthonormal rows $ \bo u_1, \ldots, \bo u_d$ that maximizes
\begin{equation}\label{eq::TSIR}
\begin{split}
\sum_{\tau \in \mathcal{T}} \left \| \diag \left( {{\bo U} \cov( \E( \bo x^{st}[t]|y[t+\tau])) {\bo U}^\top} \right) \right \|^2 = \\ \sum_{c = 1}^d \sum_{\tau \in \mathcal{T}} \left[ \bo u_c^\top \cov( \E( \bo x^{st}[t]|y[t+\tau])) \bo u_c{} \right]^2,
\end{split}
\end{equation}
for a set of chosen lags $ \mathcal{T} \subset \mathbb{Z}_ + $.
\item The value of the functional is then ${\bs \Gamma}( \bo x; y) = {\bo U {\cov( \bo x)}^{-1/2}}$.
\end{enumerate}
\end{definition}

Again, if the response time series is continuous, slicing is performed to obtain $y^{sl}[t]$ which leads to a feasible computation of the matrices $\Sigma_\tau$. However, now the optimization problem is not any more solved by a simple eigenvector-eigenvalue decomposition but joint diagonalization algorithms need to be used. There are many available options, for this paper we use the one using Given's rotations, see \cite{CardosoSouloumiac1996} for details. For other options we refer to \cite{IllnerMiettinenFuchsTaskinenNordhausenOjaTheis2015}. 
To choose the relevant vectors from the joint diagonalization \cite{MatilainenCrouxNordhausenOja2017} define
\[
\lambda_{c\tau} = \left( \bo u_c^\top \cov( \E( \bo x^{st}[t]|y[t+\tau])) \bo u_c \right)^2, \ \ \ c = 1, \ldots, p; \ \tau \in \mathcal{T}.
\]
And then choose those vectors $\bo u_c$ from the joint diagonalization where $ \lambda_{c \cdot} = \sum_{\tau \in \mathcal{T}} \lambda_{c\tau}$, $c = 1, \ldots, d$ is larger than zero.

\section{SIR for spatial data}\label{sec::sir_spatial}

In the following, as it is quite common in spatial regression, we assume that the spatial data is measured on a regular grid indexed by $[i,j]$ with $i,j \in \mathbb{Z}$.

We formulate a blind source separation model for the joint distribution of a $p$-variate explaining random field $\bo x = \bo x[i,j]$ and the univariate response random field $y=y[i,j]$ by assuming that

\[
\bo x[i,j] = \bs \Omega \bo z[i,j] + \bs \mu = \bs \Omega 
\begin{pmatrix}
\bo z^{(1)}[i,j] \\
\bo z^{(2)}[i,j]
\end{pmatrix}
+ \bs \mu.
\] 
$\bs \Omega$ and $\bs \mu$ denote the mixing matrix and location vector, as seen before.

For the latent field $\bo z = \bo z[i,j]$ we assume:

\begin{assumption}\label{model2}
	The stationary random field $\bo z[i,j]$ can be partitioned into the $d$-variate random subfield $\bo z^{(1)} = \bo z^{(1)}[i,j]$ and the $p-d$-variate random subfield $\bo z^{(2)}=\bo z^{(2)}[i,j]$. The random fields then satisfy
\begin{enumerate}
	\item[(A5)] $ \E(\bo z) = \bo 0$ and $ \cov( \bo z) = \bo I_p$, and
	\item[(A6)] $ \left(y, \bo z^{(1)}{}^\top \right)^\top  \indep \ \bo z^{(2)}$,
\end{enumerate}
where $d$ is minimal in the sense as specified in Assumption 1.
\end{assumption}
Assumption (A6) implies that $ \left(y[i,j], {\bo z^{(1)}[i,j]}^\top \right)^\top  \indep \ \bo z^{(2)}[i',j']$ which can also be expressed as $ \left(y[i-k,j-l], \bo z^{(1)}[i,j]{}^\top \right)^\top \indep \ \bo z^{(2)}[i',j']$ for all $i,i',j,j',k,l \in \mathbb{Z}$. Additionally, Assumption (A6) can also be seen as such that there exists a full rank $p \times p$ matrix $\bs \Gamma=(\bs \Gamma_1^\top,\bs \Gamma_2^\top)^\top$
such that  $\cov(\bs \Gamma x[i,j])= \bo I_p$ and 
$(y,\Gamma_1^\top \bo x)  \indep \Gamma_2^\top \bo x$.
Based on Assumption~\ref{model2} the iid sliced inverse regression (SIR) operating on the marginal distributions of $ \left(y[i,j], \bo x[i,j]^\top \right)^\top$ could also be used to identify $ \bo z^{(1)}$ but would use only the cross-sectional information while ignoring spatial dependencies. A relevant source of information we would not like to ignore. Using the idea of \cite{BeckerFried2003} and adding the neighbouring cells as additional variables to $\bo x[i,j]$ is of course again possible but even more costly. For example assuming a squared $n \times n$ grid and including just all directly connected neighbours increases the dimension eightfold and reduces the sample size to a $n^* \times n^*$ grid with $n^* = n-2$, meaning $4n-4$ observations 
are discarded. 

Like in the time series case this model formulation does not separate between independent and dependent explaining fields to explain the $y[i,j]$ field. All the dependence between the $\bo x$ and $y$ fields, as a whole, goes through $ \bo z^{(1)}$, and the aim is simply to separate between the signal field $ \bo z^{(1)}$ and the noise field $ \bo z^{(2)}$.

Again there are indeterminacies in this model formulation, the fields $ \bo z^{(1)}$ and $ \bo z^{(2)}$ are identifiable only up to pre-multiplication by orthogonal matrices. 

We define the matrices 
\[
\bs \Sigma_{(k,l)}(\bo x[i,j]) = \cov(\E(\bo x[i,j]|y[i-k,j-l])),
\]
for all $k$ $\in \mathbb{Z}$ and $l$ $\in \mathbb{Z}$ and the following holds

\begin{result}
For all random fields fulfilling (A5) and (A6)
\[
\Sigma_{(k,l)}(\bo z[i,j])
= \left(
\begin{array}{cc}
\cov \left( \E( \bm z^{(1)}[i,j]|y[i-k,j-l]) \right) & 0 \\
0 & 0 \\
\end{array}
\right),
\]
for all $k, l \in \mathbb{Z}$.
\end{result}

Finally, we have all we need to define SIR in a spatial data context which is denoted as SSIR.

\begin{definition}
	The SSIR functional ${\bs \Gamma}( \bo x;y)$ for a stationary random field $(y, \bo x^\top)^\top$ is obtained as follows.
	\begin{enumerate}
		\item Standardize $ \bo x$ and write ${\bo x}^{st} : = {\cov( \bo x)}^{-1/2}({\bo x} - \E( \bo x))$.
		\item Find the $d \times p$ matrix $ \bo U = ( \bo u_1, \ldots, \bo u_d)^\top$ with orthonormal rows $ \bo u_1, \ldots, \bo u_d$ that maximizes
		\begin{equation}\label{eq::SSIR}
\begin{split}
		\sum_{(k,l) \in \mathcal{S}} \left \| \diag \left( {{\bo U} \cov( \E( \bo x^{st}[i,j]|y[i-k, j-l])) {\bo U}^\top} \right) \right \|^2 = \\
		 \sum_{c = 1}^d \sum_{(k,l) \in \mathcal{S}} \left[ \bo u_c^\top \cov( \E( \bo x^{st}[i,j]|y[i-k, j-l])) \bo u_c{} \right]^2,
		\end{split}
\end{equation}
		for a set of chosen spatial lags $ \mathcal{S}=\{(k_1,l_1),\ldots, (k_K, l_K)\} $.
		\item The value of the functional is then ${\bs \Gamma}( \bm x;y) = {\bo U {\cov( \bm x)}^{-1/2}}$.
	\end{enumerate}
\end{definition}

Again, if the response field is continuous slicing is required for a practical computation of the inverse regression matrices $\bs \Sigma_{(k,l)}$. Here there are $K$ pairs of spatial lags as specified in the set $\mathcal{S}$, thus the solution to the maximization problem can be obtained by joint diagonalization of these former matrices.

To extract the correct directions we define in analogy to TSIR
\[
\lambda_{c(k,l)} = \left( \bo u_c^\top \cov( \E( \bo x^{st}[i,j]|y[i-k,j-l])) \bo u_c \right)^2, \ \ \ c = 1, \ldots, p; \ (k,l) \in \mathcal{S}.
\]

If the slicing did not cause any loss of information, then there should be $d$ values
 $\lambda_{c (\cdot,\cdot)} = \sum_{(k,l) \in \mathcal{S}} \lambda_{c(k,l)}$ larger 
than zero. In practise therefore these sums can give an idea about the number of directions to retain, they could be used for example in a scree plot. Furthermore the ${\lambda_{c(k,l)}}$-values contain also information about the spatial lags which might be of relevance, non-zero values indicate that the corresponding spatial lag and direction are of interest. The problem here is just if the spatial correlation in a field is large, the dependence between successive $\lambda_{c(k,l)}$ might not vanish quickly. 
To make the different $\lambda_{c(k,l)}$-values more comparable we in the following assume that the values are standardized such that they all together add up to one.

Following \cite{MatilainenCrouxNordhausenOja2017} a first option to select a number of directions and spatial lags of interest is as follows. Fix a proportion $P \in (0,1)$ and determine the smallest number of descending sorted $\lambda_{c(k,l)}$-values where the cumulative sum exceeds $P$. The lags and directions belonging to these values are then selected as directions and spatial lags of relevance. This can be compared to the strategy in principal component analysis which retains the smallest number of components that explain $100 \cdot P$\% of the variation of the data. Just here it should explain the dependence with respect to the response field $y$.  \cite{MatilainenCrouxNordhausenOja2017} also discuss other strategies to select the number of directions and  lags in a time series context based on the $\lambda_{c(k,l)}$-values. However, this approach seems to be the most promising one so far and therefore we focus on this strategy.

\begin{figure}
	\centering
	\begin{minipage}[t]{0.32\textwidth}
		\centering
		\includegraphics[width=0.95\textwidth]{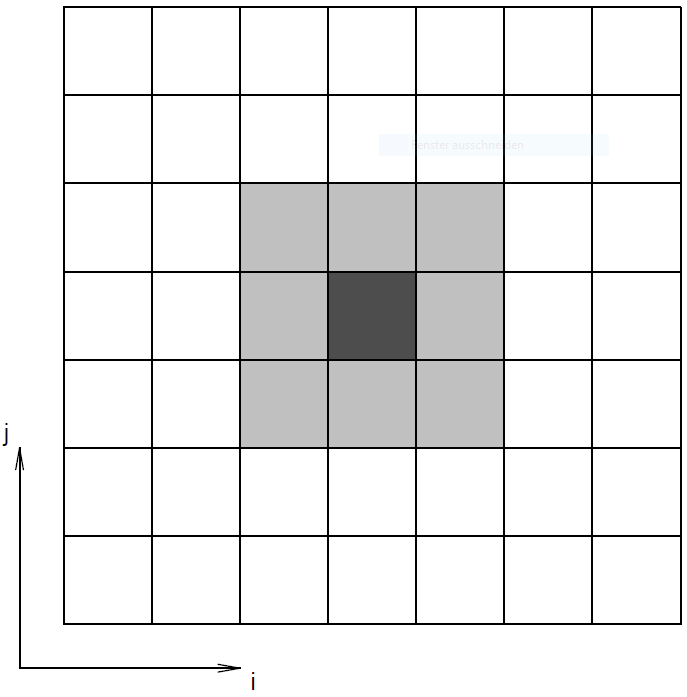}
	\end{minipage}
	\hfill
	\begin{minipage}[t]{0.32\textwidth}
		\centering
		\includegraphics[width=0.95\textwidth]{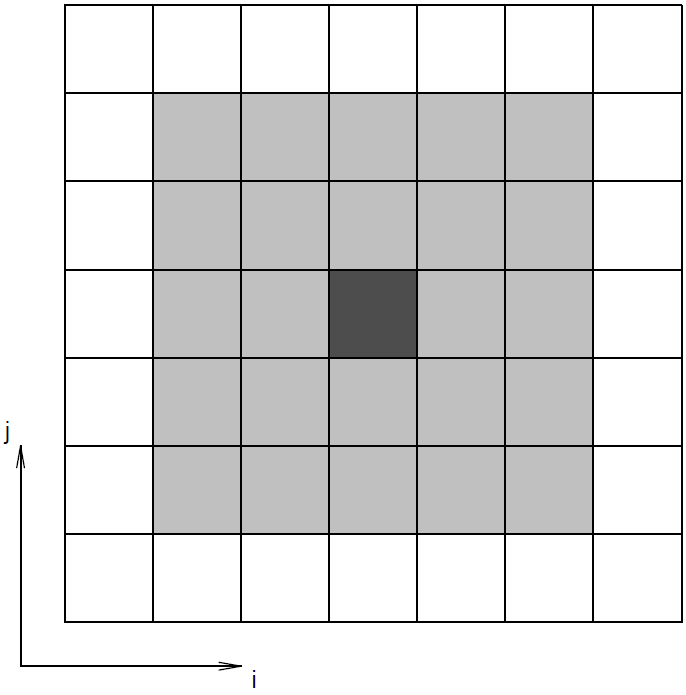}
	\end{minipage}
	\hfill
	\begin{minipage}[t]{0.32\textwidth}
		\centering
		\includegraphics[width=0.95\textwidth]{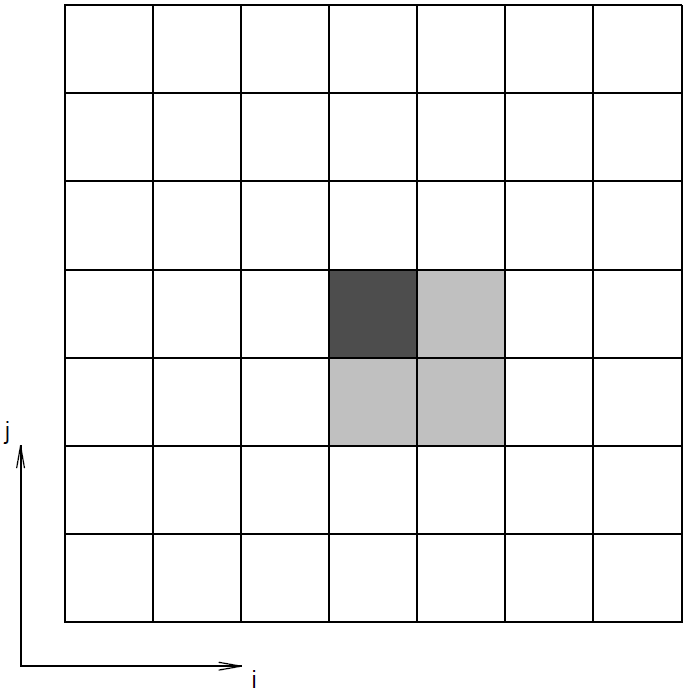}
	\end{minipage}
	\caption{Illustration of different neighbourhood relationships depicted in light grey relative to the dark grey cells. Left: Neighbours of first order. Middle: Neighbours of first and second order. Right: South-East neighbours.}
	\label{fig::grids}
\end{figure}

Naturally, the above strategies require that all relevant spatial lags are included in the set $\mathcal{S}$. In practise this should be based on expert knowledge. But in doubt the choice of lags can be very versatile. One could for example have an isotropic model in mind and include neighbours of first or neighbours of first as well as second order. In contrast, if one models data were the response is believed to be direction dependent on the predictors, then only lags of certain directions can be considered. Examples would be data where the response is dependent on wind direction. Figure~\ref{fig::grids} visualizes these three different options - in the left panel only neighbours of first order are considered which are highlighted by the light grey cells, the middle grid shows neighbours of first and second order and the right panel shows a structure based on South-East neighbours of first order. Generally, increasing the number of spatial lags leads to a higher number of $\Sigma_{(k,l)}$ matrices to be jointly diagonalized. Although theoretically adding matrices with no information has no impact, in a finite sample setting this increases the computational burden and adds noise to the joint diagonalization algorithm. Our approach of using the spatially lagged inverse regression curves is what distinguished our approach from the spatial SDR methods for grid data as described in \cite{LoubesYao2009,2019arXiv190909996A} which concentrate only on on-site information.

Another parameter needed to be chosen in SSIR is the number of slices $H$ used to estimate the inverse regression curve. We follow the usual SIR guidelines and use $H=10$ in the following simulations. This is a common choice for iid SIR which is shown to be a quite robust choice \cite{Li1991}. Similarly, for TSIR \cite{MatilainenCrouxNordhausenOja2019} carried out an extensive simulation study on the influence of $H$ on the prediction power comparing values of $H=2,5,10,20,40$. It was found that using more than $H=10$ slices does not provide meaningful results and that a lower number might be more appropriate when sample size decreases. In case of a nominal response the number of slices is naturally limited to the number of classes. For example, for a binary response variable only $H=2$ slices can be used.

\section{Performance evaluation of SSIR}\label{sec::sim}

In this section we want to evaluate how well SSIR performs. For that purpose we adapt the  three models used in a time series context in \cite{MatilainenCrouxNordhausenOja2017} to the spatial setting.

The three models are:
\begin{itemize}
	\item[\bf{A:}] 
	\quad $y[i,j] = 2 z_1 [i+1,j] + 3 z_2 [i+1,j] + \epsilon [i,j]$,
	\item[\bf{B:}] 
	\quad $y[i,j] = 2 z_1 [i+1,j] + 3 z_2 [i+2,j-2] + \epsilon [i,j]$,
	\item[\bf{C:}] 
	\quad $y[i,j] = z_1 [i+1,j] / (0.5 + ( z_2 [i+1,j] + 1.5))^2 + \epsilon [i,j]$.
\end{itemize}

Thus $\bo z^{(1)}[i,j] = \left(z_1 [i,j], \ z_2 [i,j] \right)^\top$ and we define also a two-dimensional noise part $\bo z^{(2)}[i,j] = \left(z_3 [i,j], \ z_4 [i,j]\right)^\top$. For simplicity we choose all four random fields $z_1,\ldots,z_4$ as mutually independent Gaussian random fields having mean zero and the following isotropic and homogeneous exponential covariance function  
\[
C(h) = C_0 \exp(-h/h_0) + C_1 {\bf 1}_0(h),
\]
with $C_0 = C_1 = 1$, ${\bf 1}$ is the indicator function and $h$ denotes the distance between the points under consideration. The right part in the covariance function above represents a nugget effect which is an on-site variance term, for details see \cite{TheoryOfSpatialStatistics}.

\begin{figure}
  \centering
   \includegraphics[width=0.6\textwidth]{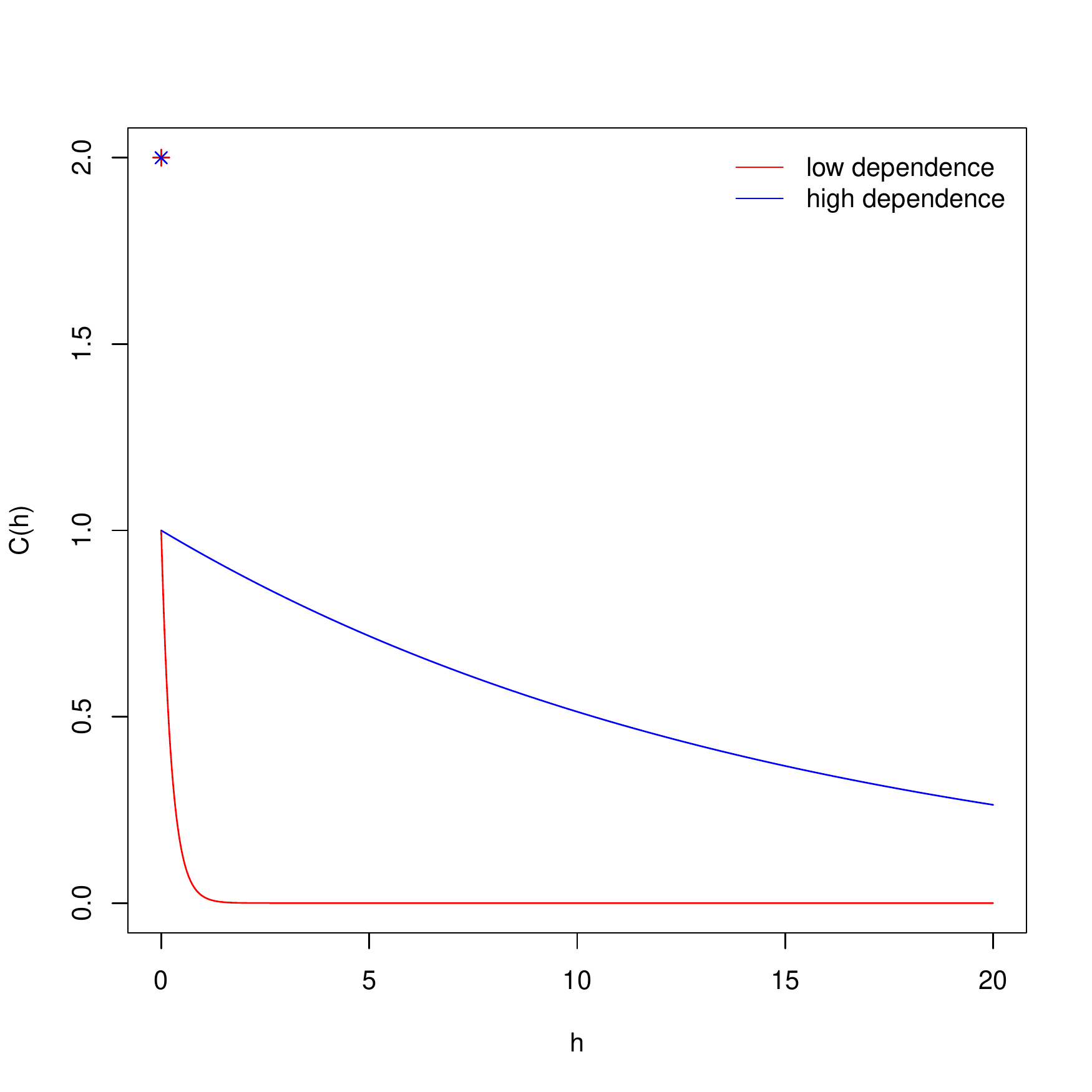}
   \caption{Visualization of the exponential covariance functions used to simulate the random fields. Choosing $h_0=0.25$ yields the weak dependence setting and $h_0=15$ the strong dependence setting.}
   \label{fig::cov_mat}
\end{figure}

In the following simulations we sample from such a field using a square grid of length 100 having 400 spatial locations in each direction yielding a total of $400\times 400$ observations. This can be thought of as having a square of 100 $\times$ 100 meters and taking a measurement every 25 centimetres. Similar as in \cite{MatilainenCrouxNordhausenOja2017} we consider two cases, in the first case there is strong spatial dependence within each of the four fields and in the second case the spatial dependence is weak. For that purpose we fix the scale parameter $h_0=0.25$ for the weak spatial dependence case and $h_0=15$ for the strong spatial dependence case. Figure~\ref{fig::cov_mat} visualizes the two different covariance functions. The discontinuity at zero represents the nugget effect. In the weak dependence case there is basically no spatial dependence when measurements are taken three units apart, whereas in the strong dependence case there is even still considerable dependence when measurements are 20 units apart. Additionally, for the independent error $\epsilon [i,j]$ we add Gaussian white noise with $\sigma^2=1$ in all three models.

As our SSIR method is affine equivariant we use without loss of generality $\bs \Omega = \bo I_4$ as the mixing matrix. Thus $\bo x[i,j] = \bo z[i,j]$ and the directions of interest in the three different models are:
\begin{itemize}
	\item[\bf{A:}] 
	\quad $((2,3,0,0)^\top \bo{x})_{[i+1,j]}$
	\item[\bf{B:}] 
	\quad $((2,0,0,0)^\top \bo{x})_{[i+1,j]}$ and $((0,3,0,0)^\top \bo{x})_{[i+2,j-2]}$
	\item[\bf{C:}] 
	\quad $((1,0,0,0)^\top \bo{x})_{[i+1,j]}$ and $((0,1,0,0)^\top \bo{x})_{[i+1,j]}$.
\end{itemize}

To fit SSIR we use either all first order neighbours or all first and second order neighbours as visualized in Figure~\ref{fig::grids} and set the number of slices $H$ to 10. 

For the computation of the following simulations we use R 3.5.1 (\cite{r_language}) with  the packages JADE (\cite{JADE_package}), raster (\cite{raster_package}), RandomFields (\cite{RandomFields_package}) and LDRTools (\cite{LDRTools_package}).

In the simulation we would like to evaluate how well SSIR estimates the directions of interest. For that purpose we consider the case where $d$ is known and also when $d$ is estimated by applying the rule described above using $P=0.5$ and $P=0.8$ respectively.
Thus, when estimating $\bs \Gamma$ the rank of the matrix might differ from the true rank $d$, this, together with the indeterminacy that the results might be rotated by an orthogonal matrix must be considered when choosing the performance criterion.

Therefore, we do not compare $\bs \Gamma$ and $\hat{\bs \Gamma}$ but their projection matrices $\bo P_{\bs \Gamma}$ and $\bo P_{\hat{\bs \Gamma}}$. Following \cite{LiskiNordhausenOjaRuizGazen2016} we measure the distance between the projection matrices using the Frobenius norm after weighting them by their rank, i.e.
\[
D_w^2(\bo P_{\hat{\bs \Gamma}}, \bo P_{{\bs \Gamma}}) = \frac{1}{2} \|w\left(\hat d\right) \bo P_{\hat{\bs \Gamma}} - w(d) \bo P_{{\bs \Gamma}} \|^2,
\] 
where for the weights we consider the two weight functions: inverse: $w(d) = 1/d$ or inverse sqrt: $w(d) = 1/\sqrt d$. Both weight function options ensure that projectors of different ranks are more comparable, they differ by their image set and interpretation in special cases, as described in \cite{LiskiNordhausenOjaRuizGazen2016}. The following results are only presented for the inverse weight function as the qualitative results are equal for both weight function options.
 
\begin{figure}
  \centering
   \includegraphics[width=0.6\textwidth]{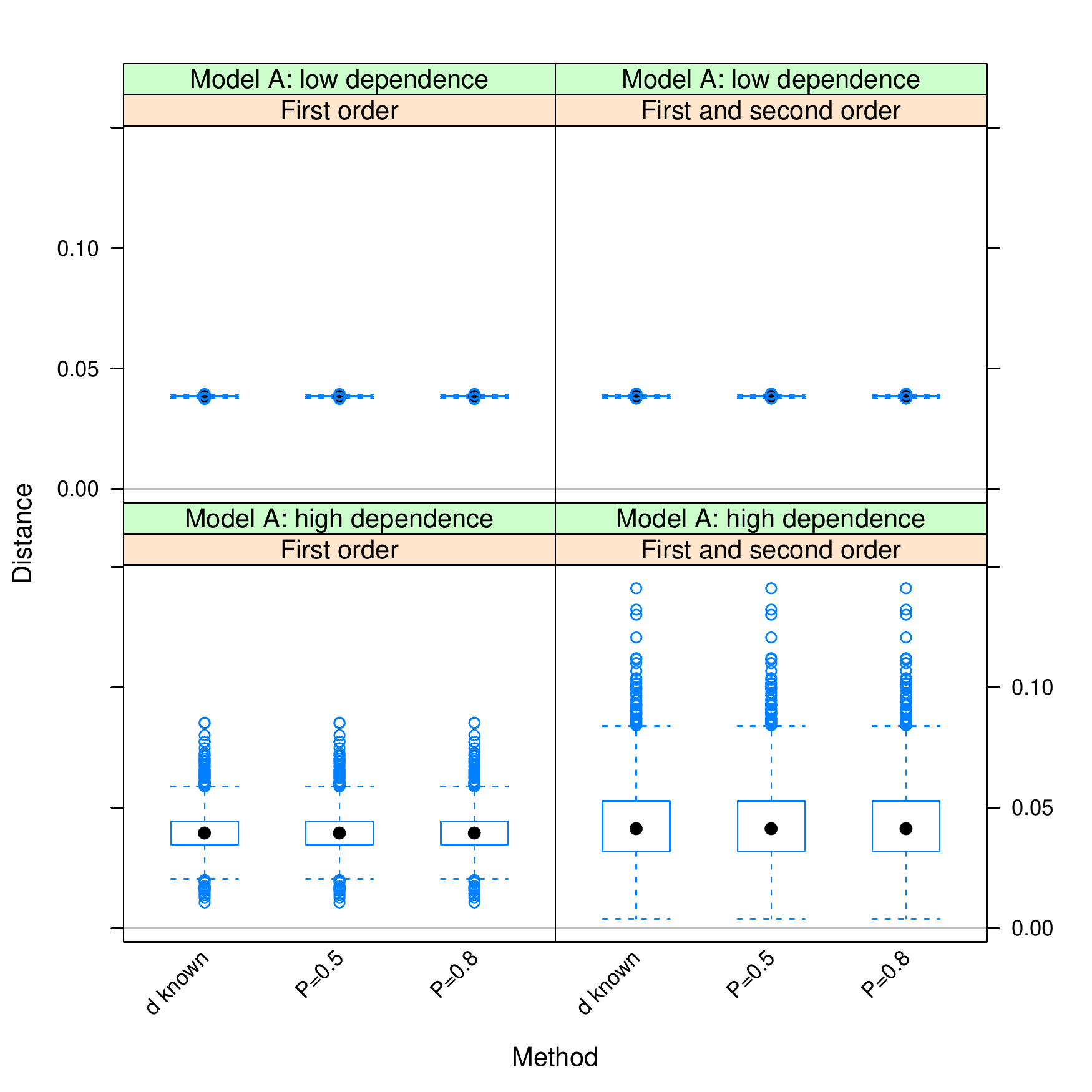}
   \caption{Inverse weighted deviations of the estimated directions projector matrices and the true directions projector matrices. 2000 repetition of Model A for the low and high dependence case are presented. Neighbours of first as well as first and second order are considered.}
   \label{fig::boxplots_A}
\end{figure}

\begin{figure}
  \centering
   \includegraphics[width=0.6\textwidth]{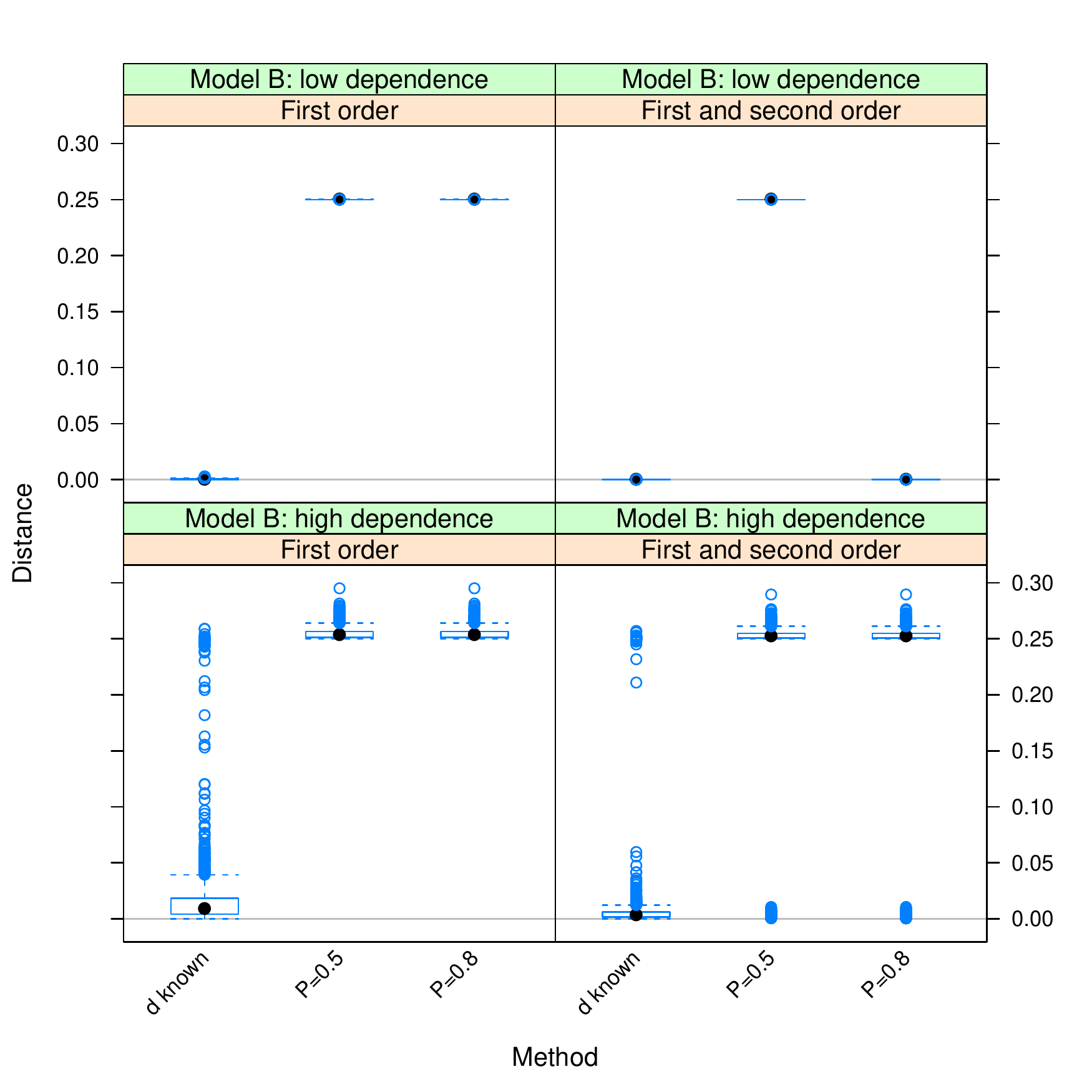}
   \caption{Inverse weighted deviations of the estimated directions projector matrices and the true directions projector matrices. 2000 repetition of Model B for the low and high dependence case are presented. Neighbours of first as well as first and second order are considered.}
   \label{fig::boxplots_B}
\end{figure}

\begin{figure}
  \centering
   \includegraphics[width=0.6\textwidth]{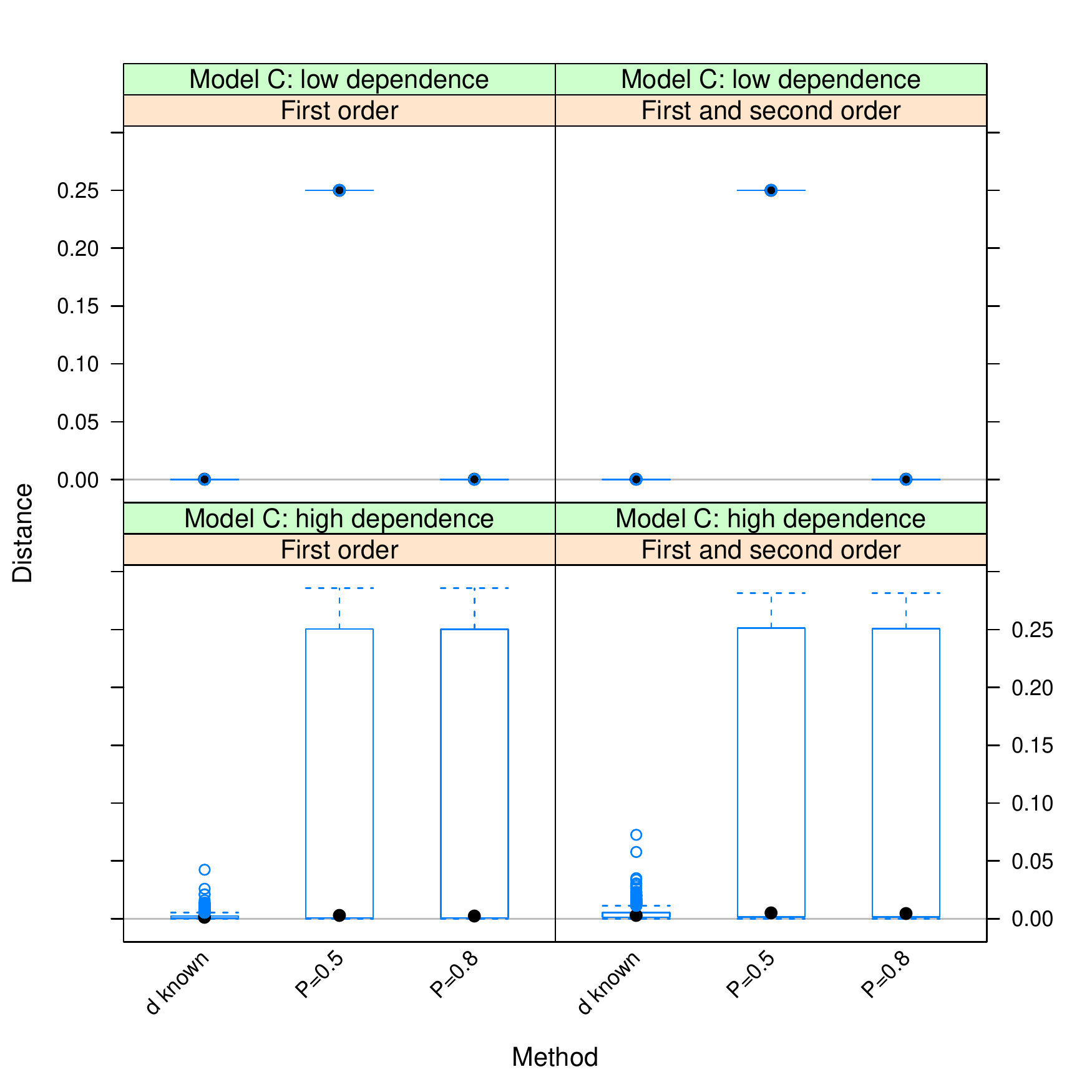}
   \caption{Inverse weighted deviations of the estimated directions projector matrices and the true directions projector matrices. 2000 repetition of Model C for the low and high dependence case are presented. Neighbours of first as well as first and second order are considered.}
   \label{fig::boxplots_C}
\end{figure}

\begin{figure}
  \centering
   \includegraphics[width=0.7\textwidth]{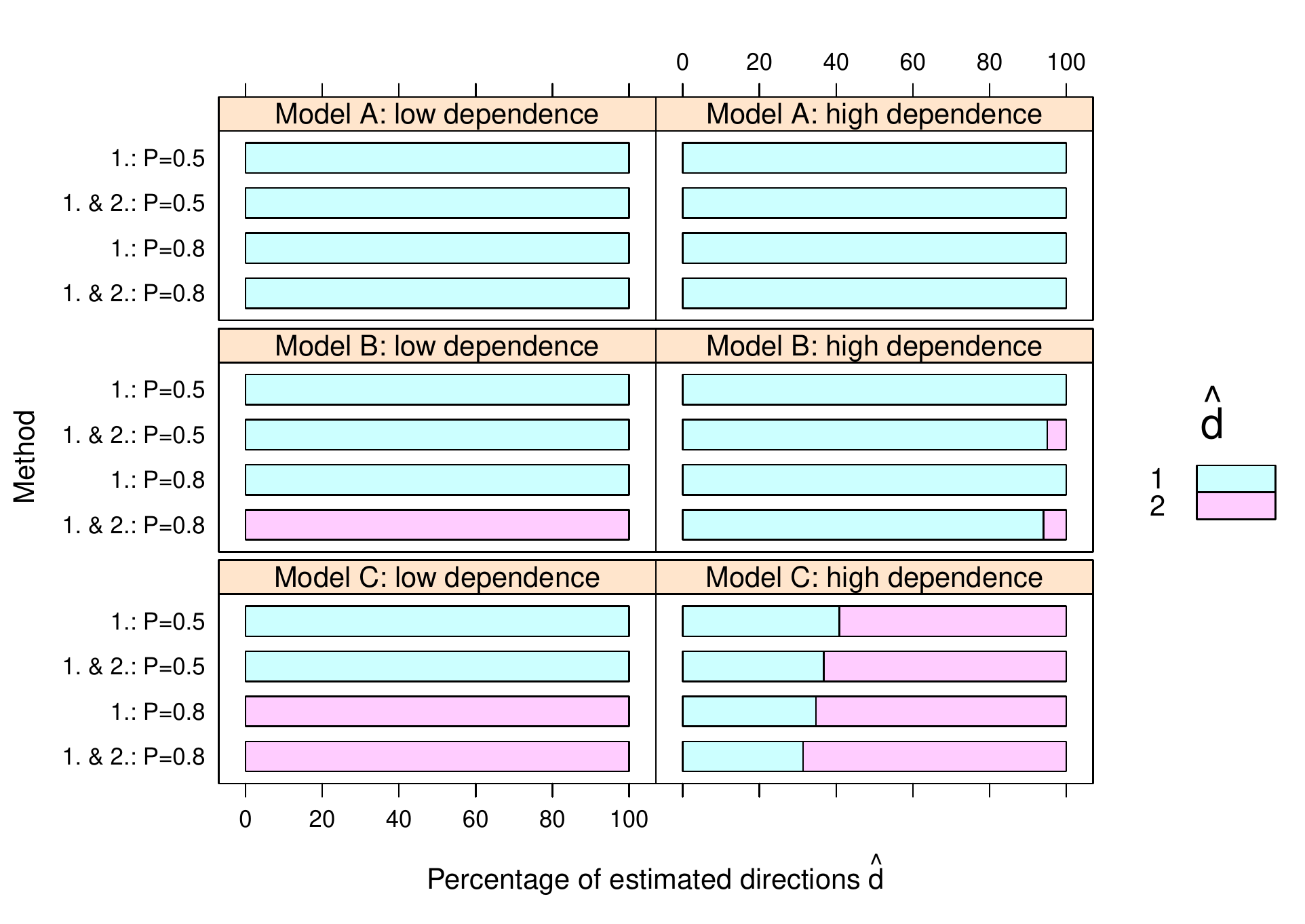}
   \caption{Percentage of estimated directions $\hat{d}$ for the simulations presented in Figures~\ref{fig::boxplots_A}-\ref{fig::boxplots_C} for different values of $P$ when using first and first and second order neighbours.}
   \label{fig::dir_d}
\end{figure}

Figures~\ref{fig::boxplots_A}-\ref{fig::boxplots_C} show the distances for all three Models in the low and high dependence settings considering first as well as first and second order neighbours for the inverse weight function based on 2000 repetitions. Furthermore, Figure~\ref{fig::dir_d} depicts the percentages of chosen directions $d$ for $P=0.5$ and $P=0.8$ of all model and dependence settings.

The distances clearly show that SSIR works as expected when the true number of directions is known as the distance to the true subspace is small. If the dimension is unknown, then the rule here using $P=0.5$ is not advisable in Models B and C. Also for Model B and C with low dependence, $P=0.8$ seems to work well except for model B when considering only first order neighbours. This effect is explained by the fact that the response in model B depends on second order neighbours as well, hence the true second direction cannot be found. If the dependence in the field is large then the performance of SSIR worsens, especially in model B and in about half of the simulations in model C directions are missed.

The problem of missing directions in the high dependence case is linked to the fact that the $\lambda_{c(k,l)}$ values do not vanish quickly and therefore also the spatial lag selection is challenging in that case. This is also observed in Figure \ref{fig::dir_d} as the number of chosen directions never exceeds the true number $d$. To illustrate this effect further we present one case from the simulation study for each model and dependence setting.

\begin{figure}
\centering
    \begin{minipage}[t]{0.9\textwidth}
        \centering
        \includegraphics[width=0.95\textwidth]{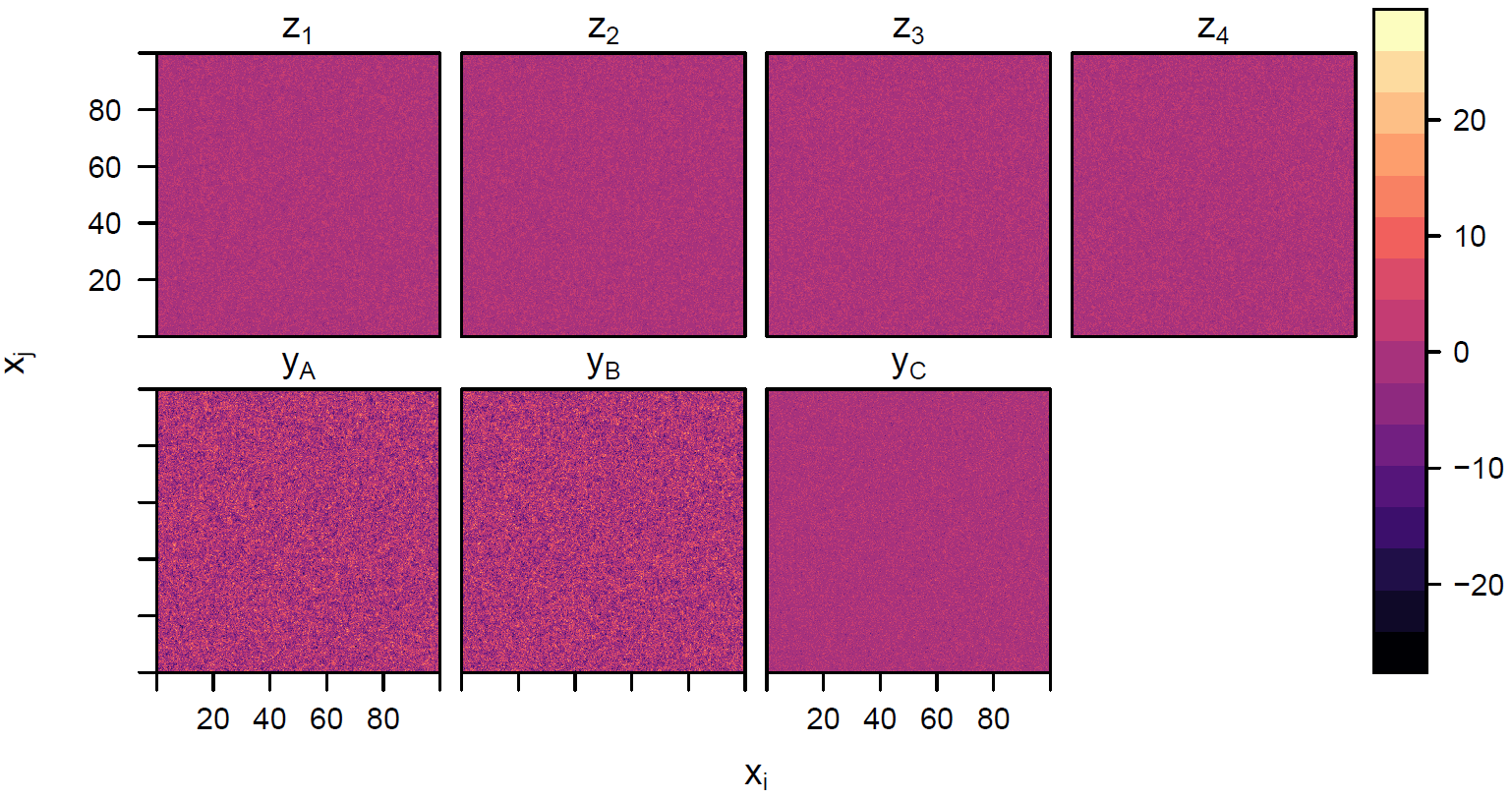}
    \end{minipage}
    \vskip\baselineskip
    \begin{minipage}[t]{0.9\textwidth}
        \centering
        \includegraphics[width=0.95\textwidth]{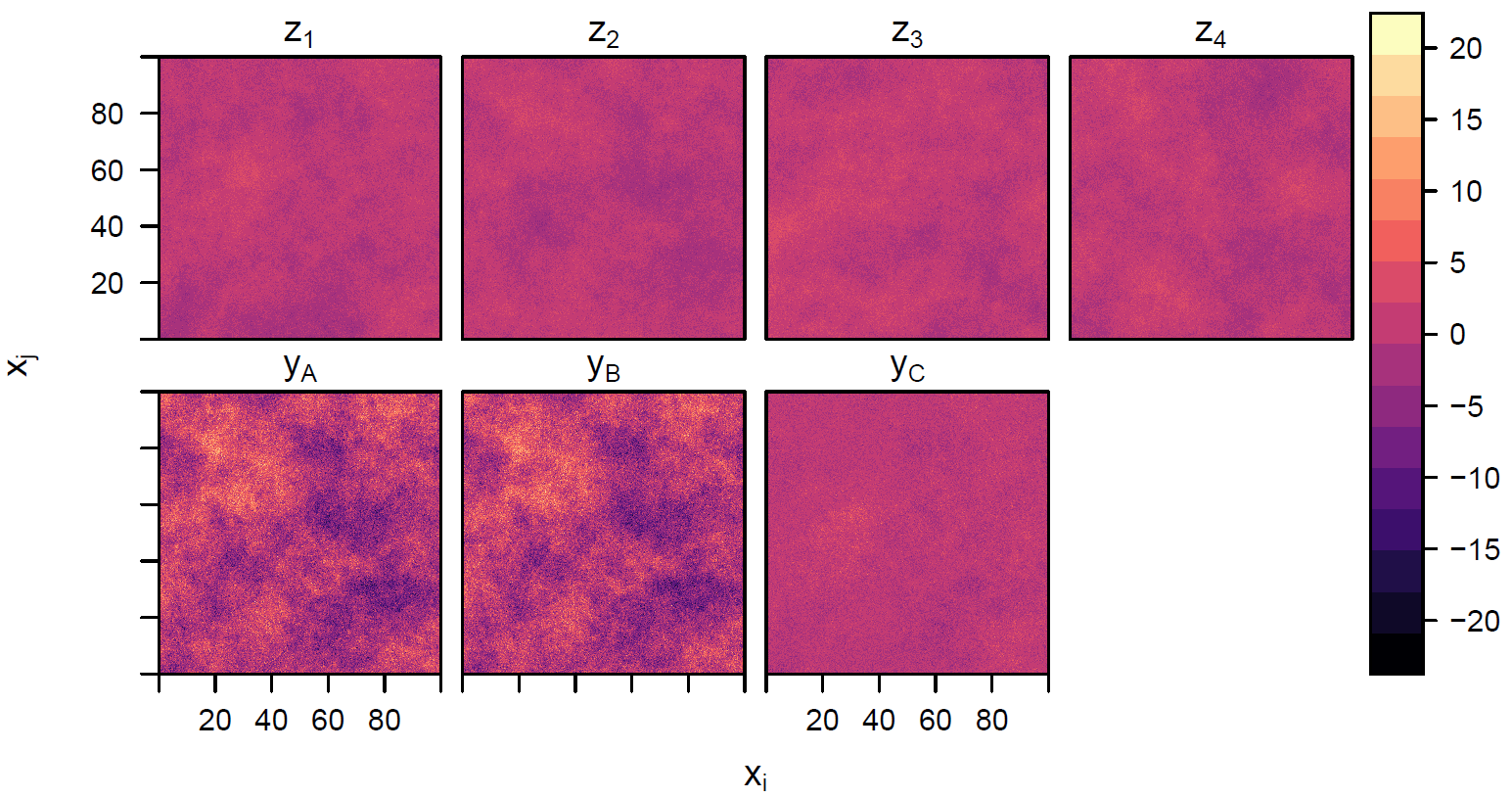}
    \end{minipage}
    \caption{Illustration of the field $\bm z$ and the responses of the three different models for low dependence (upper chart) and high dependence (lower chart).}
    \label{fig::fields}
\end{figure}

Figure~\ref{fig::fields} visualizes the latent fields and the three response fields in these 6 different settings. Tables~\ref{tab::res_A_n}-\ref{tab::res_C_nn} show the standardized $\lambda_{c(k,l)}$-values when $\mathcal{S}$ consists of all neighbours of first as well as first and second order. In the tables grey cells highlight the largest $\lambda_{c(k,l)}$-values that are needed to exceed the threshold of $P=0.8$.

\begin{table}
\centering
\small
\caption{Estimated dependencies $\hat{\lambda}_{c(k,l)}$ between $y[i,j]$ and $ (\bm u_c' \bm x^{st})_{(i+k,j+l)}$ for model A for low dependence (left table) and high dependence (right table). Considering first order neighbours. Small deviations in the sums arise from considering only four digits.}
\begin{tabular}{cc}
\hline 
\begin{minipage}{.47\linewidth}
\centering
\begin{tabular}{cc|cccc|c}
k & l & $ \bm u_1' \bm x^{st}$ & $\bm u_2' \bm x^{st}$ & $\bm u_3' \bm x^{st}$ & $\bm u_4' \bm x^{st}$ & Sum \\ 
  \hline
1 & 0 &\cellcolor[gray]{0.9} 0.9033 & 0.0001 & 0.0001 & 0.0001 & 0.9035 \\ 
  -1 & 0 & 0.0039 & 0.0001 & 0.0000 & 0.0000 & 0.0041 \\ 
  0 & 1 & 0.0134 & 0.0001 & 0.0001 & 0.0000 & 0.0136 \\ 
  0 & -1 & 0.0129 & 0.0001 & 0.0001 & 0.0001 & 0.0131 \\ 
  1 & 1 & 0.0300 & 0.0000 & 0.0000 & 0.0000 & 0.0301 \\ 
  -1 & 1 & 0.0029 & 0.0002 & 0.0000 & 0.0000 & 0.0031 \\ 
  1 & -1 & 0.0295 & 0.0001 & 0.0001 & 0.0000 & 0.0297 \\ 
  -1 & -1 & 0.0025 & 0.0000 & 0.0001 & 0.0001 & 0.0028 \\ \hline
   \multicolumn{2}{c|}{Sum} & 0.9984 & 0.0007 & 0.0005 & 0.0004 & 1 \\ 
\end{tabular}
\end{minipage} \hfill
\begin{minipage}{.47\linewidth}
\centering
\begin{tabular}{cc|cccc|c}
k & l & $ \bm u_1' \bm x^{st}$ & $\bm u_2' \bm x^{st}$ & $\bm u_3' \bm x^{st}$ & $\bm u_4' \bm x^{st}$ & Sum \\ 
  \hline
1 & 0 &\cellcolor[gray]{0.9} 0.4396 & 0.0006 & 0.0002 & 0.0001 & 0.4404 \\ 
  -1 & 0 & 0.0767 & 0.0016 & 0.0002 & 0.0000 & 0.0785 \\ 
  0 & 1 &\cellcolor[gray]{0.9} 0.0790 & 0.0015 & 0.0002 & 0.0000 & 0.0808 \\ 
  0 & -1 &\cellcolor[gray]{0.9} 0.0782 & 0.0019 & 0.0002 & 0.0001 & 0.0804 \\ 
  1 & 1 &\cellcolor[gray]{0.9} 0.0796 & 0.0015 & 0.0002 & 0.0001 & 0.0814 \\ 
  -1 & 1 &\cellcolor[gray]{0.9} 0.0770 & 0.0015 & 0.0002 & 0.0001 & 0.0788 \\ 
  1 & -1 &\cellcolor[gray]{0.9} 0.0793 & 0.0018 & 0.0003 & 0.0000 & 0.0815 \\ 
  -1 & -1 & 0.0760 & 0.0018 & 0.0002 & 0.0000 & 0.0781 \\ \hline 
  \multicolumn{2}{c|}{Sum} & 0.9853 & 0.0124 & 0.0018 & 0.0005 & 1 \\ 
\end{tabular}
\end{minipage} \\ \hline

\end{tabular}
\label{tab::res_A_n}
\end{table}

\begin{table}
\centering
\small
\caption{Estimated dependencies $\hat{\lambda}_{c(k,l)}$ between $y[i,j]$ and $ (\bm u_c' \bm x^{st})_{(i+k,j+l)}$ for model A for low dependence (left table) and high dependence (right table). Considering first and second order neighbours. Small deviations in the sums arise from considering only four digits.}
\begin{tabular}{cc}
\hline
\begin{minipage}{.47\textwidth}
\centering
\begin{tabular}{cc|cccc|c}
k & l & $ \bm u_1' \bm x^{st}$ & $\bm u_2' \bm x^{st}$ & $\bm u_3' \bm x^{st}$ & $\bm u_4' \bm x^{st}$ & Sum \\ 
  \hline
1 & 0 &\cellcolor[gray]{0.9} 0.8366 & 0.0000 & 0.0001 & 0.0000 & 0.8368 \\ 
  -1 & 0 & 0.0036 & 0.0001 & 0.0001 & 0.0000 & 0.0037 \\ 
  0 & 1 & 0.0124 & 0.0000 & 0.0001 & 0.0001 & 0.0126 \\ 
  0 & -1 & 0.0119 & 0.0001 & 0.0001 & 0.0001 & 0.0121 \\ 
  1 & 1 & 0.0278 & 0.0000 & 0.0000 & 0.0000 & 0.0279 \\ 
  -1 & 1 & 0.0026 & 0.0000 & 0.0000 & 0.0002 & 0.0029 \\ 
  1 & -1 & 0.0273 & 0.0000 & 0.0001 & 0.0001 & 0.0274 \\ 
  -1 & -1 & 0.0024 & 0.0001 & 0.0001 & 0.0000 & 0.0026 \\ 
  2 & 0 & 0.0262 & 0.0001 & 0.0000 & 0.0001 & 0.0264 \\ 
  2 & 1 & 0.0121 & 0.0001 & 0.0000 & 0.0001 & 0.0122 \\ 
  2 & 2 & 0.0025 & 0.0000 & 0.0001 & 0.0001 & 0.0027 \\ 
  2 & -1 & 0.0120 & 0.0001 & 0.0000 & 0.0000 & 0.0123 \\ 
  2 & -2 & 0.0022 & 0.0001 & 0.0001 & 0.0000 & 0.0024 \\ 
  -2 & 0 & 0.0004 & 0.0001 & 0.0001 & 0.0000 & 0.0006 \\ 
  -2 & 1 & 0.0004 & 0.0002 & 0.0001 & 0.0001 & 0.0007 \\ 
  -2 & 2 & 0.0002 & 0.0001 & 0.0001 & 0.0001 & 0.0005 \\ 
  -2 & -1 & 0.0002 & 0.0001 & 0.0001 & 0.0000 & 0.0004 \\ 
  -2 & -2 & 0.0002 & 0.0000 & 0.0000 & 0.0001 & 0.0004 \\ 
  1 & 2 & 0.0042 & 0.0001 & 0.0000 & 0.0001 & 0.0044 \\ 
  -1 & 2 & 0.0006 & 0.0002 & 0.0000 & 0.0001 & 0.0009 \\ 
  1 & -2 & 0.0041 & 0.0001 & 0.0000 & 0.0000 & 0.0042 \\ 
  -1 & -2 & 0.0009 & 0.0000 & 0.0001 & 0.0000 & 0.0011 \\ 
  0 & 2 & 0.0023 & 0.0001 & 0.0000 & 0.0001 & 0.0024 \\ 
  0 & -2 & 0.0023 & 0.0001 & 0.0001 & 0.0001 & 0.0025 \\ \hline
  \multicolumn{2}{c|}{Sum} & 0.9953 & 0.0017 & 0.0015 & 0.0015 & 1 \\ 
\end{tabular}
\end{minipage} \hfill
\begin{minipage}{.47\textwidth}
\centering
\begin{tabular}{cc|cccc|c}
k & l & $ \bm u_1' \bm x^{st}$ & $\bm u_2' \bm x^{st}$ & $\bm u_3' \bm x^{st}$ & $\bm u_4' \bm x^{st}$ & Sum \\ 
  \hline
1 & 0 &\cellcolor[gray]{0.9} 0.1952 & 0.0011 & 0.0001 & 0.0000 & 0.1964 \\ 
  -1 & 0 &\cellcolor[gray]{0.9} 0.0346 & 0.0004 & 0.0001 & 0.0000 & 0.0352 \\ 
  0 & 1 &\cellcolor[gray]{0.9} 0.0357 & 0.0004 & 0.0001 & 0.0000 & 0.0362 \\ 
  0 & -1 &\cellcolor[gray]{0.9} 0.0354 & 0.0005 & 0.0001 & 0.0000 & 0.0360 \\ 
  1 & 1 &\cellcolor[gray]{0.9} 0.0360 & 0.0004 & 0.0001 & 0.0000 & 0.0365 \\ 
  -1 & 1 &\cellcolor[gray]{0.9} 0.0348 & 0.0004 & 0.0001 & 0.0000 & 0.0353 \\ 
  1 & -1 &\cellcolor[gray]{0.9} 0.0359 & 0.0005 & 0.0001 & 0.0000 & 0.0365 \\ 
  -1 & -1 &\cellcolor[gray]{0.9} 0.0345 & 0.0005 & 0.0001 & 0.0000 & 0.0351 \\ 
  2 & 0 &\cellcolor[gray]{0.9} 0.0356 & 0.0004 & 0.0001 & 0.0000 & 0.0361 \\ 
  2 & 1 &\cellcolor[gray]{0.9} 0.0355 & 0.0004 & 0.0001 & 0.0000 & 0.0360 \\ 
  2 & 2 &\cellcolor[gray]{0.9} 0.0343 & 0.0005 & 0.0001 & 0.0000 & 0.0349 \\ 
  2 & -1 &\cellcolor[gray]{0.9} 0.0354 & 0.0005 & 0.0001 & 0.0000 & 0.0360 \\ 
  2 & -2 &\cellcolor[gray]{0.9} 0.0341 & 0.0005 & 0.0001 & 0.0000 & 0.0347 \\ 
  -2 & 0 & 0.0330 & 0.0005 & 0.0001 & 0.0000 & 0.0336 \\ 
  -2 & 1 &\cellcolor[gray]{0.9} 0.0333 & 0.0005 & 0.0001 & 0.0000 & 0.0339 \\ 
  -2 & 2 & 0.0325 & 0.0004 & 0.0001 & 0.0000 & 0.0330 \\ 
  -2 & -1 & 0.0324 & 0.0005 & 0.0001 & 0.0000 & 0.0330 \\ 
  -2 & -2 & 0.0325 & 0.0006 & 0.0001 & 0.0000 & 0.0332 \\ 
  1 & 2 &\cellcolor[gray]{0.9} 0.0350 & 0.0004 & 0.0001 & 0.0000 & 0.0355 \\ 
  -1 & 2 & 0.0332 & 0.0004 & 0.0001 & 0.0001 & 0.0337 \\ 
  1 & -2 &\cellcolor[gray]{0.9} 0.0349 & 0.0005 & 0.0001 & 0.0000 & 0.0355 \\ 
  -1 & -2 &\cellcolor[gray]{0.9} 0.0337 & 0.0005 & 0.0001 & 0.0000 & 0.0343 \\ 
  0 & 2 &\cellcolor[gray]{0.9} 0.0342 & 0.0003 & 0.0001 & 0.0001 & 0.0347 \\ 
  0 & -2 &\cellcolor[gray]{0.9} 0.0341 & 0.0006 & 0.0001 & 0.0000 & 0.0348 \\ \hline
  \multicolumn{2}{c|}{Sum} & 0.9857 & 0.0115 & 0.0021 & 0.0007 & 1 \\ 
\end{tabular}
\end{minipage} \\ \hline

\end{tabular}
\label{tab::res_A_nn}
\end{table}

\begin{table}
\centering
\small
\caption{Estimated dependencies $\hat{\lambda}_{c(k,l)}$ between $y[i,j]$ and $ (\bm u_c' \bm x^{st})_{(i+k,j+l)}$ for model B for low dependence (left table) and high dependence (right table), first order neighbours are considered. Small deviations in the sums arise from considering only four digits.}
\begin{tabular}{cc}
\hline
\begin{minipage}{.47\linewidth}
\begin{tabular}{cc|cccc|c}
k & l & $ \bm u_1' \bm x^{st}$ & $\bm u_2' \bm x^{st}$ & $\bm u_3' \bm x^{st}$ & $\bm u_4' \bm x^{st}$ & Sum \\ 
  \hline
1 & 0 &\cellcolor[gray]{0.9} 0.8687 & 0.0001 & 0.0003 & 0.0001 & 0.8692 \\ 
  -1 & 0 & 0.0049 & 0.0002 & 0.0002 & 0.0002 & 0.0055 \\ 
  0 & 1 & 0.0141 & 0.0002 & 0.0002 & 0.0001 & 0.0146 \\ 
  0 & -1 & 0.0147 & 0.0035 & 0.0002 & 0.0001 & 0.0184 \\ 
  1 & 1 & 0.0281 & 0.0005 & 0.0001 & 0.0003 & 0.0289 \\ 
  -1 & 1 & 0.0033 & 0.0003 & 0.0005 & 0.0001 & 0.0041 \\ 
  1 & -1 & 0.0322 & 0.0221 & 0.0001 & 0.0004 & 0.0547 \\ 
  -1 & -1 & 0.0033 & 0.0006 & 0.0003 & 0.0003 & 0.0044 \\ \hline 
  \multicolumn{2}{c|}{Sum} & 0.9694 & 0.0274 & 0.0017 & 0.0015 & 1 \\ 
\end{tabular}
\end{minipage} & 

\begin{minipage}{.47\linewidth}
\centering
\begin{tabular}{cc|cccc|c}
k & l & $ \bm u_1' \bm x^{st}$ & $\bm u_2' \bm x^{st}$ & $\bm u_3' \bm x^{st}$ & $\bm u_4' \bm x^{st}$ & Sum \\ 
  \hline
1 & 0 &\cellcolor[gray]{0.9} 0.2363 & 0.0153 & 0.0003 & 0.0001 & 0.2520 \\ 
  -1 & 0 &\cellcolor[gray]{0.9} 0.1016 & 0.0037 & 0.0003 & 0.0000 & 0.1055 \\ 
  0 & 1 &\cellcolor[gray]{0.9} 0.1020 & 0.0036 & 0.0002 & 0.0000 & 0.1058 \\ 
  0 & -1 &\cellcolor[gray]{0.9} 0.1049 & 0.0039 & 0.0002 & 0.0000 & 0.1091 \\ 
  1 & 1 &\cellcolor[gray]{0.9} 0.1014 & 0.0038 & 0.0003 & 0.0001 & 0.1056 \\ 
  -1 & 1 & 0.1001 & 0.0037 & 0.0003 & 0.0000 & 0.1041 \\ 
  1 & -1 &\cellcolor[gray]{0.9} 0.1070 & 0.0043 & 0.0002 & 0.0001 & 0.1116 \\ 
  -1 & -1 &\cellcolor[gray]{0.9} 0.1021 & 0.0038 & 0.0003 & 0.0001 & 0.1063 \\ \hline
  \multicolumn{2}{c|}{Sum} & 0.9554 & 0.0422 & 0.0021 & 0.0003 & 1 \\ 
\end{tabular}
\end{minipage} \\ \hline

\end{tabular}
\label{tab::res_B_n}
\end{table}

\begin{table}
\centering
\small
\caption{Estimated dependencies $\hat{\lambda}_{c(k,l)}$ between $y[i,j]$ and $ (\bm u_c' \bm x^{st})_{(i+k,j+l)}$ for model B for low dependence (left table) and high dependence (right table), first and second order neighbours are considered. Small deviations in the sums arise from considering only four digits.}
\begin{tabular}{cc}
\hline
\begin{minipage}{.47\linewidth}
\centering
\begin{tabular}{cc|cccc|c}
k & l & $ \bm u_1' \bm x^{st}$ & $\bm u_2' \bm x^{st}$ & $\bm u_3' \bm x^{st}$ & $\bm u_4' \bm x^{st}$ & Sum \\ 
  \hline
1 & 0 & 0.0026 &\cellcolor[gray]{0.9} 0.2693 & 0.0000 & 0.0001 & 0.2720 \\ 
  -1 & 0 & 0.0001 & 0.0015 & 0.0001 & 0.0000 & 0.0017 \\ 
  0 & 1 & 0.0002 & 0.0043 & 0.0000 & 0.0000 & 0.0046 \\ 
  0 & -1 & 0.0016 & 0.0041 & 0.0000 & 0.0001 & 0.0057 \\ 
  1 & 1 & 0.0004 & 0.0085 & 0.0001 & 0.0000 & 0.0091 \\ 
  -1 & 1 & 0.0001 & 0.0010 & 0.0000 & 0.0001 & 0.0013 \\ 
  1 & -1 & 0.0085 & 0.0084 & 0.0001 & 0.0000 & 0.0171 \\ 
  -1 & -1 & 0.0002 & 0.0010 & 0.0001 & 0.0001 & 0.0014 \\ 
  2 & 0 & 0.0031 & 0.0092 & 0.0001 & 0.0001 & 0.0125 \\ 
  2 & 1 & 0.0007 & 0.0039 & 0.0001 & 0.0001 & 0.0047 \\ 
  2 & 2 & 0.0002 & 0.0011 & 0.0001 & 0.0001 & 0.0014 \\ 
  2 & -1 & 0.0202 & 0.0039 & 0.0001 & 0.0001 & 0.0243 \\ 
  2 & -2 &\cellcolor[gray]{0.9} 0.6141 & 0.0003 & 0.0001 & 0.0001 & 0.6146 \\ 
  -2 & 0 & 0.0001 & 0.0003 & 0.0001 & 0.0000 & 0.0005 \\ 
  -2 & 1 & 0.0001 & 0.0002 & 0.0000 & 0.0000 & 0.0004 \\ 
  -2 & 2 & 0.0000 & 0.0001 & 0.0001 & 0.0001 & 0.0003 \\ 
  -2 & -1 & 0.0001 & 0.0002 & 0.0000 & 0.0001 & 0.0004 \\ 
  -2 & -2 & 0.0001 & 0.0002 & 0.0001 & 0.0001 & 0.0004 \\ 
  1 & 2 & 0.0001 & 0.0015 & 0.0001 & 0.0001 & 0.0018 \\ 
  -1 & 2 & 0.0000 & 0.0005 & 0.0000 & 0.0001 & 0.0006 \\ 
  1 & -2 & 0.0186 & 0.0014 & 0.0001 & 0.0000 & 0.0202 \\ 
  -1 & -2 & 0.0003 & 0.0003 & 0.0001 & 0.0000 & 0.0008 \\ 
  0 & 2 & 0.0001 & 0.0009 & 0.0001 & 0.0001 & 0.0011 \\ 
  0 & -2 & 0.0021 & 0.0011 & 0.0001 & 0.0000 & 0.0033 \\ \hline
  \multicolumn{2}{c|}{Sum} & 0.6737 & 0.3233 & 0.0016 & 0.0014 & 1 \\ 
\end{tabular}
\end{minipage} \hfill
\begin{minipage}{.47\linewidth}
\centering
\begin{tabular}{cc|cccc|c}
k & l & $ \bm u_1' \bm x^{st}$ & $\bm u_2' \bm x^{st}$ & $\bm u_3' \bm x^{st}$ & $\bm u_4' \bm x^{st}$ & Sum \\ 
  \hline
1 & 0 &\cellcolor[gray]{0.9} 0.0632 & 0.0200 & 0.0001 & 0.0001 & 0.0833 \\ 
  -1 & 0 &\cellcolor[gray]{0.9} 0.0342 & 0.0004 & 0.0002 & 0.0000 & 0.0349 \\ 
  0 & 1 &\cellcolor[gray]{0.9} 0.0345 & 0.0004 & 0.0002 & 0.0000 & 0.0351 \\ 
  0 & -1 &\cellcolor[gray]{0.9} 0.0355 & 0.0004 & 0.0002 & 0.0000 & 0.0361 \\ 
  1 & 1 &\cellcolor[gray]{0.9} 0.0344 & 0.0003 & 0.0002 & 0.0001 & 0.0350 \\ 
  -1 & 1 &\cellcolor[gray]{0.9} 0.0338 & 0.0003 & 0.0002 & 0.0001 & 0.0344 \\ 
  1 & -1 &\cellcolor[gray]{0.9} 0.0363 & 0.0003 & 0.0002 & 0.0000 & 0.0369 \\ 
  -1 & -1 &\cellcolor[gray]{0.9} 0.0344 & 0.0004 & 0.0003 & 0.0000 & 0.0352 \\ 
  2 & 0 &\cellcolor[gray]{0.9} 0.0362 & 0.0004 & 0.0002 & 0.0001 & 0.0368 \\ 
  2 & 1 &\cellcolor[gray]{0.9} 0.0351 & 0.0003 & 0.0002 & 0.0001 & 0.0356 \\ 
  2 & 2 &\cellcolor[gray]{0.9} 0.0339 & 0.0003 & 0.0002 & 0.0001 & 0.0345 \\ 
  2 & -1 &\cellcolor[gray]{0.9} 0.0369 & 0.0003 & 0.0002 & 0.0000 & 0.0374 \\ 
  2 & -2 &\cellcolor[gray]{0.9} 0.1438 & 0.0052 & 0.0010 & 0.0001 & 0.1501 \\ 
  -2 & 0 &\cellcolor[gray]{0.9} 0.0331 & 0.0004 & 0.0003 & 0.0001 & 0.0338 \\ 
  -2 & 1 & 0.0326 & 0.0003 & 0.0003 & 0.0001 & 0.0332 \\ 
  -2 & 2 & 0.0312 & 0.0003 & 0.0003 & 0.0001 & 0.0318 \\ 
  -2 & -1 & 0.0327 & 0.0004 & 0.0003 & 0.0000 & 0.0334 \\ 
  -2 & -2 &\cellcolor[gray]{0.9} 0.0331 & 0.0004 & 0.0003 & 0.0001 & 0.0338 \\ 
  1 & 2 &\cellcolor[gray]{0.9} 0.0337 & 0.0003 & 0.0002 & 0.0001 & 0.0343 \\ 
  -1 & 2 & 0.0326 & 0.0003 & 0.0002 & 0.0001 & 0.0333 \\ 
  1 & -2 &\cellcolor[gray]{0.9} 0.0361 & 0.0003 & 0.0002 & 0.0001 & 0.0367 \\ 
  -1 & -2 &\cellcolor[gray]{0.9} 0.0342 & 0.0004 & 0.0003 & 0.0000 & 0.0349 \\ 
  0 & 2 &\cellcolor[gray]{0.9} 0.0331 & 0.0003 & 0.0002 & 0.0001 & 0.0337 \\ 
  0 & -2 &\cellcolor[gray]{0.9} 0.0351 & 0.0004 & 0.0002 & 0.0000 & 0.0357 \\ \hline
  \multicolumn{2}{c|}{Sum} & 0.9596 & 0.0329 & 0.0061 & 0.0015 & 1 \\
\end{tabular}
\end{minipage} \\ \hline

\end{tabular}
\label{tab::res_B_nn}
\end{table}

\begin{table}
\centering
\small
\caption{Estimated dependencies $\hat{\lambda}_{c(k,l)}$ between $y[i,j]$ and $ (\bm u_c' \bm x^{st})_{(i+k,j+l)}$ for model C for low dependence (left table) and high dependence (right table), first order neighbours are considered. Small deviations in the sums arise from considering only four digits.}
\begin{tabular}{cc}
\hline
\begin{minipage}{.47\linewidth}
\centering
\begin{tabular}{cc|cccc|c}
k & l & $ \bm u_1' \bm x^{st}$ & $\bm u_2' \bm x^{st}$ & $\bm u_3' \bm x^{st}$ & $\bm u_4' \bm x^{st}$ & Sum \\ 
  \hline
1 & 0 &\cellcolor[gray]{0.9} 0.7106 &\cellcolor[gray]{0.9} 0.1878 & 0.0001 & 0.0001 & 0.8986 \\ 
  -1 & 0 & 0.0041 & 0.0007 & 0.0001 & 0.0002 & 0.0051 \\ 
  0 & 1 & 0.0113 & 0.0030 & 0.0002 & 0.0001 & 0.0146 \\ 
  0 & -1 & 0.0102 & 0.0038 & 0.0002 & 0.0002 & 0.0144 \\ 
  1 & 1 & 0.0229 & 0.0064 & 0.0001 & 0.0001 & 0.0296 \\ 
  -1 & 1 & 0.0027 & 0.0004 & 0.0001 & 0.0001 & 0.0032 \\ 
  1 & -1 & 0.0241 & 0.0070 & 0.0002 & 0.0001 & 0.0315 \\ 
  -1 & -1 & 0.0023 & 0.0005 & 0.0002 & 0.0001 & 0.0030 \\ \hline
  \multicolumn{2}{c|}{Sum} & 0.7883 & 0.2095 & 0.0012 & 0.0010 & 1 \\ 
\end{tabular}
\end{minipage} \hfill
\begin{minipage}{.47\linewidth}
\centering
\begin{tabular}{cc|cccc|c}
k & l & $ \bm u_1' \bm x^{st}$ & $\bm u_2' \bm x^{st}$ & $\bm u_3' \bm x^{st}$ & $\bm u_4' \bm x^{st}$ & Sum \\ 
  \hline
1 & 0 &\cellcolor[gray]{0.9} 0.3613 &\cellcolor[gray]{0.9} 0.0796 & 0.0003 & 0.0002 & 0.4413 \\ 
  -1 & 0 &\cellcolor[gray]{0.9} 0.0636 & 0.0143 & 0.0009 & 0.0003 & 0.0791 \\ 
  0 & 1 &\cellcolor[gray]{0.9} 0.0642 & 0.0156 & 0.0007 & 0.0003 & 0.0808 \\ 
  0 & -1 &\cellcolor[gray]{0.9} 0.0633 & 0.0157 & 0.0009 & 0.0002 & 0.0801 \\ 
  1 & 1 &\cellcolor[gray]{0.9} 0.0638 & 0.0148 & 0.0007 & 0.0003 & 0.0795 \\ 
  -1 & 1 &\cellcolor[gray]{0.9} 0.0630 & 0.0142 & 0.0009 & 0.0003 & 0.0784 \\ 
  1 & -1 &\cellcolor[gray]{0.9} 0.0640 & 0.0163 & 0.0012 & 0.0003 & 0.0818 \\ 
  -1 & -1 & 0.0628 & 0.0149 & 0.0009 & 0.0003 & 0.0789 \\ \hline
  \multicolumn{2}{c|}{Sum} & 0.8060 & 0.1853 & 0.0064 & 0.0023 & 1 \\ 
\end{tabular}
\end{minipage} \\ \hline

\end{tabular}
\label{tab::res_C_n}
\end{table}

\begin{table}
\centering
\small
\caption{Estimated dependencies $\hat{\lambda}_{c(k,l)}$ between $y[i,j]$ and $ (\bm u_c' \bm x^{st})_{(i+k,j+l)}$ for model C for low dependence (left table) and high dependence (right table), first and second order neighbours are considered. Small deviations in the sums arise from considering only four digits.}
\begin{tabular}{cc}
\hline
\begin{minipage}{.47\linewidth}
\centering
\begin{tabular}{cc|cccc|c}
k & l & $ \bm u_1' \bm x^{st}$ & $\bm u_2' \bm x^{st}$ & $\bm u_3' \bm x^{st}$ & $\bm u_4' \bm x^{st}$ & Sum \\ 
  \hline
1 & 0 &\cellcolor[gray]{0.9} 0.6469 &\cellcolor[gray]{0.9} 0.1711 & 0.0001 & 0.0001 & 0.8182 \\ 
  -1 & 0 & 0.0038 & 0.0006 & 0.0001 & 0.0001 & 0.0046 \\ 
  0 & 1 & 0.0104 & 0.0027 & 0.0002 & 0.0001 & 0.0133 \\ 
  0 & -1 & 0.0093 & 0.0034 & 0.0001 & 0.0001 & 0.0130 \\ 
  1 & 1 & 0.0209 & 0.0058 & 0.0001 & 0.0002 & 0.0270 \\ 
  -1 & 1 & 0.0024 & 0.0003 & 0.0001 & 0.0001 & 0.0029 \\ 
  1 & -1 & 0.0222 & 0.0064 & 0.0002 & 0.0001 & 0.0289 \\ 
  -1 & -1 & 0.0021 & 0.0004 & 0.0002 & 0.0000 & 0.0027 \\ 
  2 & 0 & 0.0235 & 0.0059 & 0.0001 & 0.0004 & 0.0299 \\ 
  2 & 1 & 0.0100 & 0.0023 & 0.0001 & 0.0002 & 0.0126 \\ 
  2 & 2 & 0.0027 & 0.0007 & 0.0002 & 0.0001 & 0.0037 \\ 
  2 & -1 & 0.0119 & 0.0021 & 0.0002 & 0.0001 & 0.0143 \\ 
  2 & -2 & 0.0024 & 0.0010 & 0.0002 & 0.0001 & 0.0037 \\ 
  -2 & 0 & 0.0006 & 0.0001 & 0.0002 & 0.0001 & 0.0010 \\ 
  -2 & 1 & 0.0006 & 0.0001 & 0.0001 & 0.0001 & 0.0009 \\ 
  -2 & 2 & 0.0002 & 0.0001 & 0.0001 & 0.0002 & 0.0005 \\ 
  -2 & -1 & 0.0004 & 0.0002 & 0.0001 & 0.0002 & 0.0008 \\ 
  -2 & -2 & 0.0005 & 0.0001 & 0.0002 & 0.0002 & 0.0010 \\ 
  1 & 2 & 0.0039 & 0.0011 & 0.0002 & 0.0001 & 0.0054 \\ 
  -1 & 2 & 0.0005 & 0.0001 & 0.0004 & 0.0002 & 0.0011 \\ 
  1 & -2 & 0.0045 & 0.0009 & 0.0001 & 0.0001 & 0.0056 \\ 
  -1 & -2 & 0.0009 & 0.0005 & 0.0002 & 0.0001 & 0.0018 \\ 
  0 & 2 & 0.0024 & 0.0007 & 0.0004 & 0.0000 & 0.0035 \\ 
  0 & -2 & 0.0022 & 0.0011 & 0.0001 & 0.0002 & 0.0036 \\ \hline
  \multicolumn{2}{c|}{Sum} & 0.7852 & 0.2078 & 0.0038 & 0.0032 & 1 \\ 
\end{tabular}
\end{minipage} \hfill
\begin{minipage}{.47\linewidth}
\centering
\begin{tabular}{cc|cccc|c}
k & l & $ \bm u_1' \bm x^{st}$ & $\bm u_2' \bm x^{st}$ & $\bm u_3' \bm x^{st}$ & $\bm u_4' \bm x^{st}$ & Sum \\ 
  \hline
1 & 0 &\cellcolor[gray]{0.9} 0.1604 &\cellcolor[gray]{0.9} 0.0352 & 0.0005 & 0.0002 & 0.1963 \\ 
  -1 & 0 &\cellcolor[gray]{0.9} 0.0286 & 0.0063 & 0.0002 & 0.0001 & 0.0353 \\ 
  0 & 1 &\cellcolor[gray]{0.9} 0.0288 & 0.0069 & 0.0002 & 0.0001 & 0.0360 \\ 
  0 & -1 &\cellcolor[gray]{0.9} 0.0284 & 0.0069 & 0.0002 & 0.0001 & 0.0356 \\ 
  1 & 1 &\cellcolor[gray]{0.9} 0.0287 & 0.0065 & 0.0002 & 0.0001 & 0.0355 \\ 
  -1 & 1 &\cellcolor[gray]{0.9} 0.0282 & 0.0063 & 0.0002 & 0.0001 & 0.0349 \\ 
  1 & -1 &\cellcolor[gray]{0.9} 0.0288 & 0.0072 & 0.0003 & 0.0001 & 0.0365 \\ 
  -1 & -1 &\cellcolor[gray]{0.9} 0.0282 & 0.0066 & 0.0003 & 0.0001 & 0.0352 \\ 
  2 & 0 &\cellcolor[gray]{0.9} 0.0293 & 0.0071 & 0.0003 & 0.0001 & 0.0367 \\ 
  2 & 1 &\cellcolor[gray]{0.9} 0.0286 & 0.0068 & 0.0002 & 0.0000 & 0.0357 \\ 
  2 & 2 &\cellcolor[gray]{0.9} 0.0283 & 0.0070 & 0.0003 & 0.0001 & 0.0357 \\ 
  2 & -1 &\cellcolor[gray]{0.9} 0.0295 & 0.0069 & 0.0003 & 0.0000 & 0.0368 \\ 
  2 & -2 &\cellcolor[gray]{0.9} 0.0277 & 0.0068 & 0.0002 & 0.0001 & 0.0348 \\ 
  -2 & 0 &\cellcolor[gray]{0.9} 0.0268 & 0.0064 & 0.0003 & 0.0001 & 0.0335 \\ 
  -2 & 1 &\cellcolor[gray]{0.9} 0.0270 & 0.0061 & 0.0002 & 0.0002 & 0.0335 \\ 
  -2 & 2 & 0.0259 & 0.0062 & 0.0002 & 0.0001 & 0.0324 \\ 
  -2 & -1 &\cellcolor[gray]{0.9} 0.0265 & 0.0064 & 0.0003 & 0.0001 & 0.0332 \\ 
  -2 & -2 &\cellcolor[gray]{0.9} 0.0272 & 0.0063 & 0.0004 & 0.0001 & 0.0339 \\ 
  1 & 2 &\cellcolor[gray]{0.9} 0.0282 & 0.0071 & 0.0002 & 0.0001 & 0.0356 \\ 
  -1 & 2 &\cellcolor[gray]{0.9} 0.0266 & 0.0064 & 0.0001 & 0.0001 & 0.0331 \\ 
  1 & -2 &\cellcolor[gray]{0.9} 0.0285 & 0.0068 & 0.0002 & 0.0001 & 0.0355 \\ 
  -1 & -2 &\cellcolor[gray]{0.9} 0.0273 & 0.0068 & 0.0003 & 0.0001 & 0.0345 \\ 
  0 & 2 &\cellcolor[gray]{0.9} 0.0281 & 0.0061 & 0.0002 & 0.0001 & 0.0345 \\ 
  0 & -2 &\cellcolor[gray]{0.9} 0.0282 & 0.0071 & 0.0002 & 0.0000 & 0.0355 \\ \hline
  \multicolumn{2}{c|}{Sum} & 0.8037 & 0.1882 & 0.0061 & 0.0021 & 1 \\ 
\end{tabular}
\end{minipage} \\ \hline

\end{tabular}
\label{tab::res_C_nn}
\end{table}

Thus the tables also show that in the low dependence setting not only the number of directions is chosen correctly but the method indicates also always the correct lags to be used. However in the high dependence setting often one direction dominates and almost all spatial lags are considered informative, this effect is again in accordance to the findings depicted in Figure~\ref{fig::dir_d}.

\section{Discussion}

In this work we suggested an extension to SIR in the spatial data setting. If the number of directions of interest is known SSIR works well. However, the identification of the important spatial lags seems to depend strongly on the dependence within the latent fields. Therefore further research is needed to help decide on the number of directions and spatial lags to be considered.

We investigated SSIR on data lying on a 2-dimensional regular gird. Clearly more dimensions can be easily added to the grid in order to consider for example rather voxels than pixels. In general the data does not need to be on a regular grid as concrete distances between locations are not required in SSIR. In many spatial econometric applications neighbour definitions are rather loose as spatial units, for example administrative districts, are used, this is no problem for SSIR either. If however the locations are irregularly measured and the distances between locations matter, then SSIR needs to be adapted. Here ideas from the recently suggested spatial blind source separation approaches in \cite{NordhausenOjaFilzmoserReimann2015,BachocGentonNordhausenRuizGazenVirta2018} could be applied by drawing rings of different radii around each location and average over them to define a neighbourhood when assuming an isotropic relationship.


\section{Acknowledgments}
The work of CM and KN was supported by the Austrian Science Fund (FWF) Grant number P31881-N32.

\bibliographystyle{unsrt}

\end{document}